\documentclass[12pt]{iopart}

\usepackage{iopams}  
\usepackage{setstack}

\usepackage{subfigure}% Include figure files
\usepackage{subeqn}% Include figure files
\usepackage{graphicx}% Include figure files

\graphicspath{{./}{./images/}{./images_all/}{./images_all/GSHE_energy_dissipation/}}
\DeclareGraphicsExtensions{.eps}
\begin{document}

\title[]{Ultra-low-energy computing paradigm using giant spin Hall devices}

\author{Kuntal Roy}
\address{School of Applied and Engineering Physics, Cornell University, Ithaca, New York 14853, USA\\ *Present Address: School of Electrical and Computer Engineering, Purdue University, West Lafayette, IN 47907, USA.}

\ead{royk@purdue.edu}

\date{\today}% It is always \today, today,

\begin{abstract}
Spin Hall effect converts charge current to spin current, which can exert spin-torque to switch the magnetization of a nanomagnet. Recently, it is shown that the ratio of spin current to charge current using spin Hall effect can be made more than unity by using the areal geometry judiciously, unlike the case of conventional spin-transfer-torque switching of nanomagnets. This can enable energy-efficient means to write a bit of information in nanomagnets. Here, we study the energy dissipation in such spin Hall devices. By solving stochastic Landau-Lifshitz-Gilbert equation of magnetization dynamics in the presence of room temperature thermal fluctuations, we show a methodology to simultaneously reduce switching delay, its variance and energy dissipation, while lateral dimensions of the spin Hall devices are scaled down. 
\end{abstract}

\maketitle

\section{Introduction}

Spintronics is a promising field of research that can possibly replace the traditional charge-based electronics~\cite{RefWorks:328,RefWorks:761,RefWorks:763}. Spin-transfer-torque (STT) is a current-induced magnetization switching (CIMS) mechanism in which magnetization of a nanomagnet can be switched between its two stable states~\cite{RefWorks:7,RefWorks:173,RefWorks:16,roy11_3} or it may lead to oscillatory motion of magnetization~\cite{RefWorks:737}, however, the energy dissipation in such switching mechanism is too high for practical application purposes compared to the traditional charge-based electronics~\cite{RefWorks:472}. There are other mechanisms that are coming along for energy-efficient switching of a bit of information e.g., electric field-induced magnetization switching in multiferroic heterostructures~\cite{roy13_spin,roy11,roy14_2,RefWorks:806,RefWorks:609}, perpendicular anisotropy~\cite{RefWorks:775,RefWorks:774,RefWorks:777}, coupled polarization-magnetization switching in single-phase multiferroic materials~\cite{RefWorks:664,RefWorks:558,RefWorks:665,RefWorks:322}. However, one mechanism that has gotten attention recently is to utilize the spin Hall effect, which was first recognized by D'yakonov and Perel'~\cite{RefWorks:760} and following which there have been both theoretical studies~\cite{RefWorks:764,RefWorks:770,RefWorks:771,RefWorks:765} and experimental investigations in semiconductors~\cite{RefWorks:807,RefWorks:766,RefWorks:767,RefWorks:769}. Utilizing spin Hall effect to generate a sufficient spin current for technological purposes of exciting magnetization dynamics was severely limited~\cite{RefWorks:759}, however, there have been recent resurgence of interests~\cite{RefWorks:759,RefWorks:772,RefWorks:810,RefWorks:811} due to \emph{giant} spin Hall effect of exerting spin-torque, which can be used to switch the magnetization direction of a nanomagnet using different spin Hall materials e.g., platinum~\cite{RefWorks:757,RefWorks:818,RefWorks:814,RefWorks:817}, tantalum~\cite{RefWorks:608}, tungsten~\cite{RefWorks:758}, CuBi~\cite{RefWorks:755}.

\begin{figure}
\includegraphics{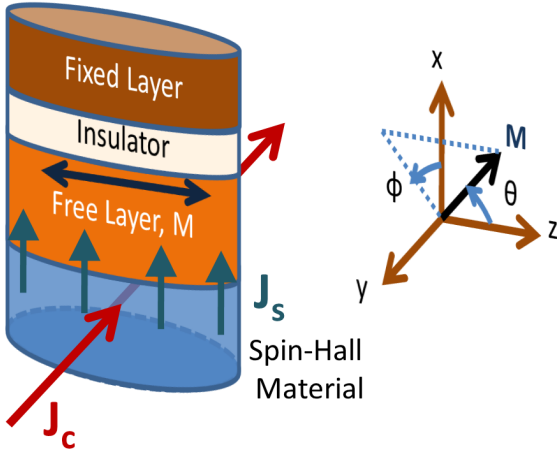}
\caption{\label{fig:GSHE_schematic} Simplified schematic diagram of a giant spin Hall device and axis assignment. The charge current (of current density $\mathbf{J_c}$) flows in the $y$-direction through the spin Hall material layer accumulating spins of opposite polarities on its opposite surfaces, which generates a spin current (of current density $\mathbf{J_s}$) in the $x$-direction with spin polarization in the $z$-direction. For the direction of charge current and spin current shown in the diagram, the direction of spin polarization at the free layer-spin Hall material layer interface will be in the $+z$-direction for materials with \emph{positive} spin Hall angle (e.g., Pt), while it will be in the $-z$-direction for materials with \emph{negative} spin Hall angle (e.g., CuBi).  Since the two mutually anti-parallel stable magnetization orientations of the free layer along the $z$-axis encode the logic bits 0 and 1, the spin-torque acting in the free layer needs to switch the magnetization in either direction and it can be performed by changing the direction of the charge current. Note that unlike the case of traditional spin-transfer-torque devices, the charge current is not spin-polarized by the fixed layer, however, the fixed layer and the insulator (e.g., MgO) are \emph{required} to \emph{read} the magnetization state of the free layer by passing a small charge current in the $x$-direction via tunneling magnetoresistance (TMR) mechanism. In standard spherical coordinate system, $\theta$ is the polar angle and $\phi$ is the azimuthal angle.}
\end{figure}

Fig.~\ref{fig:GSHE_schematic} shows a schematic diagram of a spin Hall device. A charge current flows laterally in the spin Hall material layer and a perpendicular spin current through the cross-section of the structure is generated, which can exert spin-torque on the free layer nanomagnet and switch its magnetization. In the conventional STT devices, the charge current gets spin polarized through a fixed layer, however, passing high current during switching of magnetization occasionally damages the thin insulator layer~\cite{RefWorks:608}. Hence, the giant spin Hall devices offer a great advantage over the traditional STT devices. Note that the read current, which is small, still needs to be passed perpendicular to the cross-section of the spin Hall devices and hence the fixed layer in a magnetic tunnel junction (MTJ) structure is required to detect the magnetization states of the free layer nanomagnet~\cite{RefWorks:577}. By reversing the direction of the charge current, the magnetization of the free layer can be switched in either direction. Since the charge current and the spin current flow though different cross-sections, it is shown that the spin-to-charge current ratio can be greater than one, unlike the case of traditional STT devices~\cite{RefWorks:608}. Both the Oersted field generated by the current and Rashba field are neglected since experimentally it has been verified that they tend to oppose the switching and of small magnitude~\cite{RefWorks:608}.

Here, we perform an analysis on the energy dissipation in the aforesaid spin Hall devices. Since the charge current flows through the spin Hall material, the Ohmic loss occurs in the spin Hall material itself unlike the case of traditional STT devices, where the energy dissipation incurs in the MTJ stack. The charge current flows \emph{laterally} through a higher resistance compared to the case if flown perpendicular to the cross-section (of the nanomagnet), however, the MTJ stack (nanomagnets separated by an insulator) has a higher resistance than that of the current flow path in the spin Hall material layer, particularly while using a low-resistivity material for the spin Hall layer. Also, the required charge current is less due to higher spin-to-charge conversion factor in \emph{giant} spin Hall devices. We show that the energy dissipation is inversely proportional to the lateral area of the spin Hall devices, but scaling down the lateral dimensions also makes the required spin current to switch the magnetization smaller. We solve stochastic Landau-Lifshitz-Gilbert (LLG) equation~\cite{RefWorks:162,RefWorks:161,RefWorks:186} of magnetization dynamics to depict a strategy whereby the energy dissipation can be reduced simultaneously with lowering of the switching delay and its variance in the presence of room temperature thermal fluctuations, while scaling down the lateral dimensions.

\section{Model}

We consider the free layer nanomagnet in the shape of an elliptical cylinder with its elliptical cross section lying on the $y$-$z$ plane; the major axis is along the $z$-direction and the minor axis is along the $y$-direction (see Fig.~1). The dimensions along the $z$-, $y$- , and $x$-direction are $a$, $b$, and $l$ ($a > b \gg l$), respectively. So the lateral area of the spin Hall device is $A = (\pi/4)ab$ and the nanomagnet's volume is $\Omega = (\pi/4)abl$. In standard spherical coordinate system, $\theta$  is the polar angle and the $\phi$ is the azimuthal angle.  Due to shape anisotropy, the two degenerate magnetization states along the $z$-direction ($\theta=0^\circ$ and $180^\circ$) can store a binary bit of information. The $y$-axis is the in-plane hard axis and $x$-axis is the out-of-plane hard axis. We can write the shape anisotropy energy of the nanomagnet as follows:
\begin{equation}
E_{shape}(\theta,\phi) = \frac{1}{2} \, M \, [H_k  + H_d\, cos^2\phi] \, \Omega \, sin^2\theta,
\label{eq:shape_anisotropy}
\end{equation}
where
$M=\mu_0 M_s$, $\mu_0$ is permeability of free space, $M_s$ is the saturation magnetization, $H_k=(N_{dy}-N_{dz})\,M_s$ is the Stoner-Wohlfarth switching field~\cite{RefWorks:157}, $H_d=(N_{dx}-N_{dy})\,M_s$ is the out-of-plane demagnetization field~\cite{RefWorks:157}, and $N_{dm}$ is the m$^{th}$ ($m=x,y,z$) component of the demagnetization factor~\cite{RefWorks:402} ($N_{dx} \gg N_{dy} > N_{dz}$). The shape anisotropic energy barrier between the two stable magnetization states can be expressed from Equation~(\ref{eq:shape_anisotropy}) [putting $\phi=\pm 90^\circ$] as  
\begin{equation}
E_{barrier} = \frac{1}{2} \, M \, H_k\, \Omega.
\label{eq:barrier}
\end{equation}
The magnetization \textbf{M} of the nanomagnet has a constant magnitude but a variable direction, and hence we can represent it by a vector of unit norm $\mathbf{n_m} =\mathbf{M}/|\mathbf{M}| = \mathbf{\hat{e}_r}$, where $\mathbf{\hat{e}_r}$ is the unit vector in the radial direction in spherical coordinate system represented by ($r$,$\theta$,$\phi$); the other two unit vectors in the spherical coordinate system are $\mathbf{\hat{e}_\theta}$ and $\mathbf{\hat{e}_\phi}$ for $\theta$ and $\phi$ rotations, respectively. 

The effective field and torque acting on the magnetization due to gradient of potential landscape can be expressed as $\mathbf{H_{eff}}(\theta,\phi) = - \nabla E_{shape}(\theta,\phi) = - (\partial E_{shape}/\partial \theta)\,\mathbf{\hat{e}_\theta} - (1/sin\theta)\,(\partial E_{shape}/\partial \phi)\,\mathbf{\hat{e}_\phi}$ and $\mathbf{T_E}(\theta,\phi)= \mathbf{n_m} \times \mathbf{H_{eff}}(\theta,\phi)$, respectively.
\begin{equation}
\mathbf{T_E} (\theta,\phi) = - B_{shape}(\phi)\, sin(2\theta) \,\mathbf{\hat{e}_\phi} - B_{shape,\phi}(\phi)\,sin\theta \,\mathbf{\hat{e}_\theta}, 							 
\label{eq:torque}
\end{equation}
\noindent
where 
\begin{eqnarray}
B_{shape}(\phi) &=& (1/2) \, M\, [H_k  + H_d\, cos^2\phi] \, \Omega, \\
B_{shape,\phi}(\phi) &=& (1/2) \, M H_d \, \Omega \,  sin(2\phi). 
\label{eq:B_Bshape_phi}
\end{eqnarray}

Passage of a spin current $I_s$ in the $x$-direction with spin polarization along the $z$-direction generates a spin-transfer-torque that is given by~\cite{RefWorks:7}
\begin{equation}
\mathbf{T_{STT}}(t) = - s\, \mathbf{n_m}(t) \times \left(\mathbf{n_s} \times \mathbf{n_m}(t)\right) = s  \, sin\theta \, \mathbf{\hat{e}_\theta},
\label{eq:STT_torque}
\end{equation}
\noindent
where $s = (\hbar/2e) I_s$ is the spin angular momentum deposition per unit time, the unit vector $\mathbf{n_s} = +\mathbf{\hat{e}_z}$ is in the direction of spin polarization since we want to rotate magnetization from $-z$-axis to $+z$-axis, and we have used the identity $\mathbf{\hat{e}_z} = cos\theta\,\mathbf{\hat{e}_r} - sin\theta\,\mathbf{\hat{e}_\theta}$. No asymmetry in spin-transfer-torque switching or field-like torque is considered here~\cite{RefWorks:14,RefWorks:15,RefWorks:815}.

The random thermal fluctuations are incorporated via a random magnetic field $\mathbf{h}(t)= h_x(t)\mathbf{\hat{e}_x} + h_y(t)\mathbf{\hat{e}_y} + h_z(t)\mathbf{\hat{e}_z}$, where $h_i(t)$ ($i=x,y,z$) are the three components of the random thermal field in Cartesian coordinates. We assume the same properties of the random field $\mathbf{h}(t)$ as described in Ref.~\cite{RefWorks:186}. The random thermal field can be written as~\cite{RefWorks:186}
\begin{equation}
h_i(t) = \sqrt{\frac{2 \alpha kT}{|\gamma| M \Omega \Delta t}} \; G_{(0,1)}(t) \quad (i \in x,y,z),
\label{eq:ht}
\end{equation}
\noindent
where $\alpha$ is the phenomenological dimensionless Gilbert damping parameter, $k$ is the Boltzmann constant, $T$ is temperature, $\gamma$ is the gyromagnetic ratio for electrons and its magnitude is equal to $2.21\times10^5$ (rad.m).(A.s)$^{-1}$, $\Delta t$ is the simulation time-step used, and the quantity $G_{(0,1)}(t)$ is a Gaussian distribution with zero mean and unit variance. 

The thermal field and the corresponding torque acting on the magnetization can be written as $\mathbf{H_{TH}}(\theta,\phi,t)=P_\theta(\theta,\phi,t)\,\mathbf{\hat{e}_\theta}+P_\phi(\theta,\phi,t)\,\mathbf{\hat{e}_\phi}$ and $\mathbf{T_{TH}}(\theta,\phi,t)=\mathbf{n_m} \times \mathbf{H_{TH}}(\theta,\phi,t)$, respectively, 
where
\begin{eqnarray}
P_\theta(\theta,\phi,t) &=& M  \lbrack h_x(t)\,cos\theta\,cos\phi + h_y(t)\,cos\theta sin\phi - h_z(t)\,sin\theta \rbrack \,\Omega,\\
P_\phi(\theta,\phi,t) &=& M  \lbrack h_y(t)\,cos\phi -h_x(t)\,sin\phi\rbrack \,\Omega.
\label{eq:thermal_parts}
\end{eqnarray}

The magnetization dynamics under the action of the torques $\mathbf{T_{E}}(t)$, $\mathbf{T_{STT}}(t)$, and 
$\mathbf{T_{TH}}(t)$ is described by the stochastic Landau-Lifshitz-Gilbert (LLG) equation~\cite{RefWorks:162,RefWorks:161,RefWorks:186} as follows.
\begin{equation}
\frac{d\mathbf{n_m}}{dt} - \alpha \left(\mathbf{n_m} \times \frac{d\mathbf{n_m}}{dt} \right) = -\frac{|\gamma|}{M\Omega} \left\lbrack \mathbf{T_E}+ \mathbf{T_{STT}} +  \mathbf{T_{TH}}\right\rbrack.
\end{equation}

From the above equation, we get the following coupled equations of magnetization dynamics for $\theta$ and $\phi$:
\begin{eqnarray}
\left(1+\alpha^2 \right) \frac{d\theta}{dt} &=& \frac{|\gamma|}{M \Omega} [ (B_{shape,\phi}(\phi) - s)\,sin\theta \nonumber\\
																						&& - \alpha B_{shape}(\phi) sin(2\theta) + \left(\alpha P_\theta(\theta,\phi,t) + P_\phi (\theta,\phi,t) \right) ],
 \label{eq:theta_dynamics}
\end{eqnarray}
\begin{eqnarray}
\left(1+\alpha^2 \right) \frac{d \phi}{dt} &=& \frac{|\gamma|}{M \Omega} [ \alpha (B_{shape,\phi}(\phi) - s)  + 2 B(\phi) cos\theta \nonumber\\
																					&&- {[sin\theta]^{-1}} \left(P_\theta (\theta,\phi,t) - \alpha P_\phi (\theta,\phi,t) \right) ] \;
	(sin\theta \neq 0).
\label{eq:phi_dynamics}
\end{eqnarray}
We solve the above two coupled equations numerically to track the trajectory of magnetization over time.

The energy dissipated in the nanomagnet due to Gilbert damping can be expressed as  $E_{damp} = \int_0^{\tau}P_{damp}(t) dt$, where $\tau$ is the switching delay and $P_{damp}(t)$ is the power dissipated at time $t$ given by
\begin{equation}
P_{damp}(t) = \frac{\alpha \, |\gamma|}{(1+\alpha^2) M \Omega} \, |\mathbf{T_E} (\theta(t), \phi(t)) + \mathbf{T_{STT}}(t)|^2 .
\label{eq:power_dissipation}
\end{equation}
Thermal field with zero mean does not cause any net energy dissipation but it causes variability in the energy dissipation by scuttling the trajectory of magnetization.

\begin{figure*}
\includegraphics{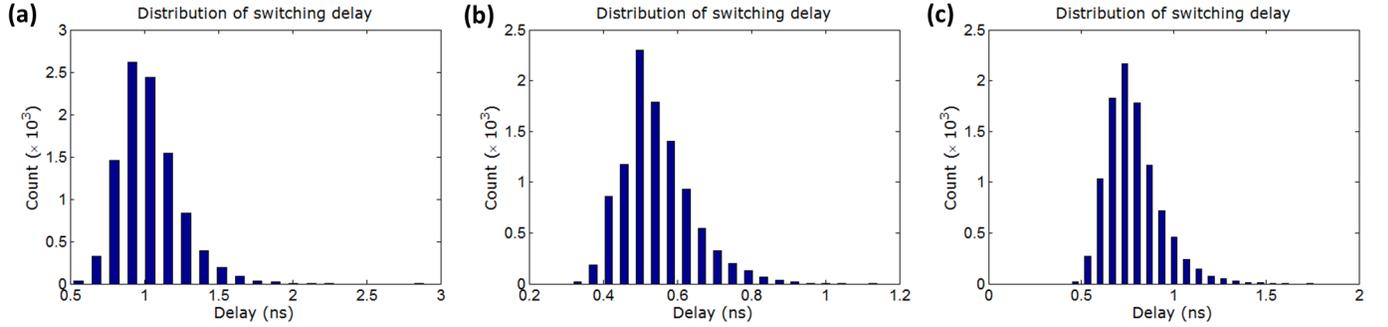}
\caption{\label{fig:switching_delay_distributions} Switching delay distributions from 10,000 simulations by solving stochastic Landau-Lifshitz-Gilbert (LLG) equation of magnetization dynamics in the presence of room temperature (300 K) thermal fluctuations. (a) $a$=150 nm, $b$=100 nm, $l$=2 nm, and $I_c$ = 133.3 $\mu$A. The mean of switching delay distribution and its standard deviation are 1.03 ns and 0.21 ns, respectively. (b) $a$=100 nm, $b$=30 nm, $l$=2 nm, and $I_c$ = 198.8 $\mu$A. The mean of switching delay distribution and its standard deviation are 0.55 ns and 0.10 ns, respectively. (c) $a$=100 nm, $b$=30 nm, $l$=2 nm, and $I_c$ = 133.3 $\mu$A. The mean of switching delay distribution and its standard deviation are 0.78 ns and 0.15 ns, respectively. The distributions look like log-normal distributions and can be fitted accordingly. The means and the standard deviations are directly calculated from the numerically computed distributions.}
\end{figure*}

If the magnetization situates {\it exactly} along the easy axis, i.e., $\sin\theta=0$ ($\theta=0^\circ$ or $\theta=180^\circ$), the torque acting on the magnetization given by Equations~(\ref{eq:torque}) and~(\ref{eq:STT_torque}) becomes zero. That is why only thermal fluctuations can deflect the magnetization vector \emph{exactly} from the easy axis. Magnetization fluctuates around an easy axis due to thermal agitations and hence we get a distribution of the initial angles ($\theta_{initial},\phi_{initial}$). We consider this initial distribution when magnetization starts switching from $\theta \simeq 180^\circ$. We perform a moderately large number (10,000) of simulations in the presence of thermal fluctuations and when the final value $\theta_{final}$ becomes $\leq 4.5^\circ$ (note that the mean value of $\theta_{initial}$ is $\sim$$4.5^\circ$), the switching is deemed to have completed. Then, the mean and standard deviation of switching delay distribution ($\tau_{mean}$ and $\tau_{std}$, respectively), and the mean of energy dissipation $E_{damp,mean}$ are extracted from the simulations. (See Fig.~\ref{fig:switching_delay_distributions} and Table~\ref{tab:Ed_comparison} later.)

The energy dissipation due to Gilbert damping is of the order of the energy barrier height, however, the major part of the energy dissipation occurs due to Ohmic loss in the spin Hall material since the charge current $I_c$ flows through it in the $y$-direction (see Fig.~\ref{fig:GSHE_schematic}). The spin-current $I_s$ flows in the $x$-direction, and the ratio of the spin-to-charge current can be expressed as 
\begin{equation}
\frac{I_s}{I_c} = \frac{\mathbf{J_s}}{\mathbf{J_c}} \, \frac{\mathbf{A_s}}{\mathbf{A_c}} = \Theta_{SH}\,\frac{b}{t_{SH}},
\label{eq:GSHE}
\end{equation}
where $\mathbf{J_s}$ ($\mathbf{J_c}$) is the spin (charge) current density, $\mathbf{A_s}$ ($\mathbf{A_c}$) is the area through which spin (charge) current flows, $\Theta_{SH}$ is the spin Hall angle of the spin Hall material used, and $t_{SH}$ is the thickness of the spin Hall material layer. The geometric factor $b/t_{SH}$ can be selected much greater than one and hence, having $\Theta_{SH}\,b > t_{SH}$ can make $I_s > I_c$.

The power dissipation in the spin Hall material can be expressed as
\begin{eqnarray}
P_d = I_c^2 R &=& \left(\frac{t_{SH}I_s}{\Theta_{SH} b}\right)^2 \left(\rho\, \frac{2}{\pi} \,\frac{b}{a\,t_{SH}} \right) \nonumber\\
							&=& \frac{1}{2}\, \left(\frac{\rho}{\Theta_{SH}^2}\right) \, \left(\frac{t_{SH}}{A}\right)\,I_s^2,
\label{eq:power}
\end{eqnarray}
where $\rho$ is the resistivity of the spin Hall material and $R$ is the resistance of the spin Hall material layer. The energy dissipation in the spin Hall material layer can be expressed as
\begin{equation}
E_d = P_d T_p = \frac{1}{2}\, \left(\frac{\rho}{\Theta_{SH}^2}\right) \, \left(\frac{t_{SH}}{A}\right)\,I_s^2 \, T_p,
\label{eq:energy}
\end{equation}
where $T_p$ is the time-period until which the charge current is kept active.

\section{Results}

For the free layer nanomagnet, we consider the widely-used material CoFeB, which has low saturation magnetization $M_s = 8 \times 10^5$ A/m~\cite{RefWorks:413} and a low Gilbert damping parameter $\alpha$ = 0.01. Note that the damping parameter can be modified depending on the adjacent spin Hall material layer due to spin pumping~\cite{RefWorks:816}. We choose CuBi as the spin Hall material since it has the lowest $\rho/\Theta_{SH}^2$ factor ($\rho$=10 $\mu\Omega$-$cm$,  $\Theta_{SH}$=--0.24 measured at 10 K~\cite{RefWorks:755}) among the other materials utilized in current experiments with an eye to reduce the energy dissipation [see Equation~(\ref{eq:energy})]. We can also possibly utilize platinum as the spin Hall material layer that has $\rho \simeq 15$ $\mu\Omega$-$cm$~\cite{RefWorks:818},  $\Theta_{SH}$=0.07~\cite{RefWorks:817} measured at room temperature, however, it will incur an energy dissipation of one order more than that of CuBi. Note that platinum increases magnetization damping (i.e., increases the \emph{minimum} switching current) of CoFeB considerably and the damping parameter is about three times compared to when tantalum is utilized~\cite{RefWorks:608}.

Note that CoFeB has a resistivity of 100 $\mu\Omega$-$cm$~\cite{RefWorks:784}, which is 10 times more than that of CuBi spin Hall material layer. Hence the current shunting effect~\cite{RefWorks:782,RefWorks:785} through the CoFeB layer is ignored. Therefore, the purpose of utilizing a low-resistive spin Hall material layer is also to tackle the issue of current shunting effect, apart from reducing the energy dissipation. We do need to consider the current shunting effect while utilizing tantalum or tungsten as the spin Hall material layer since they have the resistivity of the same order as of CoFeB~\cite{RefWorks:608,RefWorks:758}. We consider here in-plane switching of magnetization, however, perpendicular switching of magnetization has been demonstrated~\cite{RefWorks:774,RefWorks:819,RefWorks:815,RefWorks:608} and can be considered too, particularly to scale down the lateral area of the devices further.

We use both the thicknesses $t_{SH}$ and $l$ as 2 nm. We vary the lateral dimensions $a$ and $b$ of the nanomagnet for different cases considered and we choose the dimensions of the nanomagnet so that it has a single ferromagnetic domain~\cite{RefWorks:133,RefWorks:402}. We always keep the energy barrier $E_{barrier}$ = 0.8 eV or 32 $kT$ at room temperature ($T$=300 K). Following Boltzmann distribution, this means that the error probability due to spontaneous reversal of magnetization is $e^{-E_{barrier}/kT} = e^{-32}$ or $10^{-14}$, which is low enough for application purposes. We can choose a higher barrier height leading to even lower error probability depending on the application requirements. We will now go through the following three cases.

%\begin{tabular*}{l c c c c c c c c c c}
%\begin{tabular*}{\textwidth}{@{}l*{15}{@{\extracolsep{0pt plus12pt}}l}}

\begin{table}[t]
\caption{\label{tab:Ed_comparison} Different metrics for three different cases considered. Cases (a) and (b) correspond to the same dissipated power $P_d$= 377 nW, while the cases (a) and (c) correspond to the same charge current $I_c$ = 133.3 $\mu$A. The energy dissipation is reduced by about 5 times for the case (c) compared to the case (a), while the lateral area is scaled down by 5 times.}
\footnotesize\rm
\begin{tabular*}{\textwidth}{@{}c c c c c c c c c c c}
\br
Case & Nanomagnet size  & $I_s$  & $I_c$ & $R$ & $\tau_{mean}$ & $\tau_{std}$ & $E_{damp,mean}$ & $T_p$ & $P_d$ & $E_d$ \\
     & (nm$^3$) & (mA) & (mA) & ($\Omega$) & (ns) & (ns) & (aJ) & (ns) & (nW) & (aJ)\\
\mr
(a) & 150 $\times$ 100 $\times$ 2 & 1.6000 & 0.1333 & 21.22 & 1.03 & 0.21 & 0.0883 & 2.5 & 377 & 943\\
(b) & 100 $\times$ 30 $\times$ 2  & 0.7155 & 0.1988 & 9.55  & 0.55 & 0.10 & 0.0395 & 1.2 & 377 & 453\\
(c) & 100 $\times$ 30 $\times$ 2  & 0.4800 & 0.1333 & 9.55	& 0.78 & 0.15 & 0.0485 & 2.0 & 170 & 204\\
\br
\end{tabular*}
\end{table}

\textit{Case (a):} We consider a nanomagnet with dimensions $a$ = 150 nm, $b$ = 100 nm, and $l$ = 2 nm. We solve stochastic LLG for 10,000 times and find that it requires about $I_s$ = 1.6 mA to switch always successfully, while the switching current is kept active for 2.5 ns. The corresponding charge current is $I_c$ = 133.3 $\mu$A. The demagnetization factor ($N_{dx},\,N_{dy},\,N_{dz}$) = (0.9468, 0.0339, 0.019)~\cite{RefWorks:402}, $H_k=0.0148$ T, and $H_d=0.9177$ T. The switching delay distribution is plotted in the Fig.~\ref{fig:switching_delay_distributions}(a). The mean of switching delay and its standard deviation are 1.03 ns and 0.21 ns, respectively. The major energy dissipation is due to Ohmic loss in the spin Hall material and it turns out to be 943 aJ [see Table~\ref{tab:Ed_comparison}].

\textit{Case (b):} We now reduce the lateral dimensions of the nanomagnet and choose the lateral dimensions to be $a$ = 100 nm, $b$ = 30 nm with an eye to increase the device density on a chip. The thickness $l$ is kept same as for the case (a). Hence, both the lateral area and the volume decrease by a factor of 5. Note that the energy barrier between the two stable states is kept constant [see Equation~(\ref{eq:barrier})] at this reduced volume by choosing the elliptical cross-section more anisotropic. The demagnetization factor ($N_{dx},\,N_{dy},\,N_{dz}$) = (0.8829, 0.0953, 0.0218)~\cite{RefWorks:402}, $H_k=0.0739$ T, and $H_d=0.7917$ T. The anisotropic field $H_k$ increases due to the modification of demagnetization factors at the chosen dimensions~\cite{RefWorks:402}. However, at a reduced volume, the magnetization becomes more prone to thermal fluctuations [see Equation~(\ref{eq:ht})]. But the STT switching current also mitigates the detrimental effects of thermal agitations. Since in this case the current needs to switch a nanomagnet of a smaller volume [see Equations~(\ref{eq:theta_dynamics}) and~(\ref{eq:phi_dynamics})], we show that it is possible to adjust the switching current such that in overall, we can achieve a \emph{better performance metrics for both switching delay and energy dissipation at this reduced volume}. 

According to the Equation~(\ref{eq:power}), we scale down $I_s$ by a factor of $\sqrt{5}$ to keep the power dissipation constant. The corresponding distribution of switching delay is shown in the Fig.~\ref{fig:switching_delay_distributions}(b). Note that both the mean and standard deviation of switching delay have got reduced by half compared to the case (a). The switching current is kept on until 1.2 ns and hence the energy dissipation is reduced by more than half compared to the case (a). Note that even if the spin current $I_s$ is reduced, the charge current required in this case has got increased compared to the case (a) due to the decrease in minor axis $b$ of the elliptical cross-section of the nanomagnet [see Equation~(\ref{eq:GSHE})]. Next we consider another case where the spin current $I_s$ is reduced further to keep the charge current same compared to the case (a).

\textit{Case (c):} We choose the same dimensions of the nanomagnet as of the case (b) and the same charge current as for the case (a) [see Table~\ref{tab:Ed_comparison}]. The demagnetization factor, $H_k$, and $H_d$ are same as of the previous case. The corresponding switching delay distribution is plotted in the Fig.~\ref{fig:switching_delay_distributions}(c). Note that both the mean and standard deviation of switching delay distribution have got increased compared to the case (b) due to decrease in $I_s$, however, the metrics are still much better than that of the case (a) [see Table~\ref{tab:Ed_comparison}]. The energy dissipation $E_d$ has got reduced by more than half, compared to the case (b), to 204 aJ, which is due to the decrease in spin current $I_s$.

\bigskip
It should be noted that it is not only the \emph{mean} of switching delay distribution, but also the \emph{standard deviation} in switching delay that plays an important role in setting the clock period of magnetization switching for application purposes. A higher standard deviation would lead to setting a higher clock period. In this paper, it is shown that both the \emph{mean} and the \emph{standard deviation} of switching delay can be reduced \emph{simultaneously}, while the lateral area of the giant spin Hall devices is scaled down.

\section{Discussions}

From Equation~\eref{eq:GSHE}, it should be noted that the spin diffusion length $\lambda_{SH}$ for spin Hall material is not considered in the expression. For a thick spin Hall material layer ($t_{SH} \gg \lambda_{SH}$), it does not need to consider $\lambda_{SH}$. However, we should choose the thickness of the spin Hall material layer $t_{SH}$ small since the charge current (for a required spin current) decreases with the decrease of $t_{SH}$. It is possible to add a spin-sink layer at the bottom of a \emph{thin} spin Hall material layer (see Fig.~\ref{fig:GSHE_schematic}) to avoid any backflow of spins. For example, see Ref.~\cite{RefWorks:804} for some experimental results, however, research on such front is quite emerging. For CuBi utilized as a spin Hall material layer, such experimental data is not available, however, this concept of adding a spin-sink layer is quite general. Also it should be noted that there is controversy on the experimentally measured spin diffusion length for spin Hall materials~\cite{RefWorks:804}. Since for all the three cases in Table~\ref{tab:Ed_comparison}, the results correspond to the same thickness of the spin Hall material layer, the \emph{comparative} nature of the analysis is not quite affected.  The analysis presented here depicts the necessity of decreasing the thickness of the spin Hall material layer to decrease the charge current and consequently to reduce the energy dissipation.

It is imperative to compare the energy savings utilizing the \emph{giant} spin Hall effect compared to the traditional way of exerting spin-torque. The energy savings accrue from the decrease in charge current due to the geometric factor and decrease in resistance of the charge flow path compared to the MTJ stack. The reduction of energy dissipation is as high as 3-4 orders of magnitude compared to the conventional spin-transfer-torque switching mechanism~\cite{RefWorks:296,RefWorks:786} and domain-wall racetrack memory~\cite{RefWorks:501,RefWorks:329}.

Although the energy dissipation in these giant spin Hall devices has got reduced to the order of 0.1 fJ, the energy dissipation can be further reduced by having a spin Hall material that has even lower $\rho/\Theta_{SH}^2$ factor [see Equation~(\ref{eq:energy})], i.e., having a material with a lower resistivity and a higher spin Hall angle. The target would be the reduction of energy dissipation by a factor of 2 more to be competitive with other emerging technologies~\cite{roy13_spin}. Switching delay and area of a device also need to be competitive with the traditional transistor based technology~\cite{itrs}. The non-volatility of the nanomagnets in these giant spin Hall devices can be utilized to devise a possibly better architecture in terms of performance metrics for application purposes.

\section{Conclusions}

In conclusion, we have analyzed the energy dissipation in recently proposed spin Hall devices exploiting giant spin Hall effect. After formulating the energy dissipation, we solved stochastic Landau-Lifshitz-Gilbert equation of magnetization dynamics in the presence of room-temperature thermal fluctuations to present a methodology in which both the energy dissipation and switching delay (and its variance) can be reduced simultaneously, while the lateral dimensions of the spin Hall devices are scaled down. The energy dissipation turns out to be of several orders of magnitude less than that of the traditional spin-transfer-torque devices. This field is still emerging and with suitable spin Hall materials, the energy dissipation can be reduced further. This opens up an energy-efficient avenue to control a bit of information in nanomagnetic memory and logic systems for our future information processing paradigm.

\section*{References}
%\bibliographystyle{iopart-num}
%\bibliography{royk,royk2}
\providecommand{\newblock}{}

\end{document}